# LYMAN $\alpha$ FORESTS
# AND COOLING OUTFLOW FROM DWARF GALAXIES

Boqi Wang

Department of Physics and Astronomy
The Johns Hopkins University, Baltimore, MD 21218

# LYMAN $\alpha$ FORESTS AND COOLING OUTFLOW FROM DWARF GALAXIES

Boqi Wang[*]
Department of Physics and Astronomy
The Johns Hopkins University, Baltimore, MD 21218

## ABSTRACT

We propose that cooling outflows from star-forming dwarf galaxies that are spatially correlated with but unbound to bright galaxies may account for the low column QSO Ly$\alpha$ forests, providing an explanation for both the heavy elements recently detected in them and the large sizes associated with them. Thus, we suggest that structure formations in the early universe on scales where matters collapse gravitationally and the gas forms stars are responsible for the Ly$\alpha$ forests (cf. the mini-halo model). Supernovae shock the interstellar gas to high temperature, and the resulting gas outflows from the dwarfs owing to their shallow potential wells. The gas cools radiatively as the radiative cooling time shortens because of adiabatic expansion. Subsequent thermal instability results in condensation of clouds. The clouds inherit the kinetic energy of the flow and coast to large distances, providing regions of the absorbing gas with sizes of up to hundreds of kpc. We calculate the Ly$\alpha$ absorption line profiles and the neutral column distribution, and show that they are consistent with observations. We suggest that the Ly$\alpha$ forests are caused by the faint blue galaxies found in the deep extragalactic surveys.
*Subject Headings:* quasars: absorption lines–galaxies: evolution–galaxies: distances and redshifts

## 1. INTRODUCTION

Recent studies of the QSO Ly$\alpha$ forests suggest that they may be associated with galaxies. The spectroscopy and broad-band imaging of galaxies toward 3C 273 indicate that the Ly$\alpha$ absorbers are not distributed at random with respect to the galaxies (Morris et al. 1993). The initial results of an imaging and spectroscopic survey of the HST target QSOs suggest that the Ly$\alpha$ absorptions are associated with the galaxies (Lanzetta et al. 1995). Furthermore, spectroscopic studies of pairs of QSOs at high redshifts show that the median radius of the Ly$\alpha$ clouds is 100-300$h^{-1}$ kpc ($h$ is the Hubble constant $H_0$ in units of 100 km/s/Mpc) (Bechtold et al. 1994; Dinshaw et al. 1994). Indeed, recent detection of heavy elements in Ly$\alpha$ clouds (Tytler 1995) implies that star formation is involved.

Behcall & Spitzer (1969) first suggested that quasar absorption lines are caused by the gas associated with normal galaxies. Wang (1993) attributed the origin of the high column (the heavy element and Lyman limit) absorption lines to gas accretion by satellite galaxies bound to the normal galaxies. A way of confining gas clouds in the halos is by thermal pressure as suggested by Mo (1994). In this *Letter*, however, we propose that the low-column Ly$\alpha$ forests result from cooling outflows from star-forming dwarf galaxies that are spatially correlated with (but unbound to) normal galaxies. Supernovae (SN) shock heat the interstellar gas, producing a hot phase ($T \sim 10^6$ K), independent of the galaxy mass (e.g., Cox & Smith 1974; McKee & Ostriker 1977). While for normal, massive, galaxies like our own the hot gas is likely to be confined in the lower galactic halo, it is most likely to outflow out of dwarf galaxies because of their shallow potential wells (Wang 1993; 1995).

## 2. COOLING OUTFLOWS FROM DWARFS

We consider radial (spherically symmetric), steady, outflows starting from an initial radius ($r_i$) with an initial temperature ($T_i$). Given $r_i$, $T_i$, and the mass loss rate $\lambda$ (or equivalently the initial gas density), the outflow solutions are completely specified. The general solutions of the outflow problem have been obtained in Wang (1995), and we refer that paper for details. In the present work we focus on supersonic outflows from dwarf galaxies where gas can easily escape. For clarity, below we discuss a specific example to illustrate characteristics of the outflows.

---

[*] Also Space Telescope Science Institute

We adopt the cooling function for a gas with 1/100 solar abundance from Boringer & Hensler (1989) (also see Sutherland & Dopita 1993), consistent with the inferred abundance of the Ly$\alpha$ forests (e.g, Lu 1991; Tytler & Fan 1994; Tytler 1995). The cooling is not sensitive to the abundance for such a low-abundance gas since hydrogen and helium recombination lines become important. Note that because each SN typically shocks $\sim 10^4$ M$_\odot$ of the interstellar gas (e.g., McKee 1989), the abundance of the hot gas will not be much affected by the heavy elements ejected from the present SN (average $\sim 0.2$ M$_\odot$; Wang & Silk 1993) unless the gas is almost primordial. The above abundance is also roughly compatible with that observed in the most of nearby dwarf irregulars (Skillman et al. 1989). We ignore photoelectric heating by the diffuse UV background (not important at high temperatures), but assume that the gas is kept at $2 \times 10^4$ K by photoelectric heating once it cools below it.

The local dwarf spheroids are shown to contain substantial dark matter, and those with adequate kinematic data appear to have essentially flat velocity dispersion profiles (e.g., Mateo 1994). HI synthesis studies of the blue compact dwarf galaxies show extended HI disk much larger than the optical counterpart, with approximately flat rotation curves at outer regions (Meurer 1994). We thus consider dwarf galaxies whose dominant dark matter is distributed as isothermal spheres but truncated at radius $r_t$; we take $r_t = 10$ kpc below (e.g., Meurer 1994). The condition for the gas to cool radiatively is not influenced by $r_t$, neither is the fate of the cooled clouds so long as the cloud velocity is larger than the escape velocity (Wang 1995). We take the circular (rotation) velocity of the dark matter halo to be $v_{cir} = 50$ km/s. If a strong photoionizing background could inhibit the formation of galaxies with comparable $v_{cir}$ (Efstathiou 1992), the adopted $v_{cir}$ should be larger (e.g., $v_{cir} = 100$ km/s). But the basic results below should not be affected by the change because the clouds can still escape the galaxy.

We show in figure 1 the resulting supersonic outflow with $r_i = 2$ kpc, $T_i = 4 \times 10^5$ K and $\lambda = 0.5$ M$_\odot$/yr. The flow is driven by thermal pressure, and the gas first cools adiabatically. As the temperature $T$ reaches about $10^5$ K, the radiative cooling time ($t_c$) is equal to the flow time ($t_f = r/v$), so radiative cooling starts dominating. At $T \simeq 2 \times 10^4$ K, $t_c$ is only a small fraction of $t_f$ and the cooling is totally dominated by radiative loss. In the above temperature range, the gas is thermally unstable (Field 1965), and consequently, we expect that the gas cools to condense into clouds. Following Wang (1995), we assume that cloud formation completes at cooling radius $r_c = 7$ kpc where $T$ drops to $2 \times 10^4$ K. The cooled clouds inherit the kinetic energy with velocity $v_c = 168$ km/s, well beyond the escape velocity for the galaxy. Therefore, they coast away from $r_c$ and eventually leave the galaxy, making a large region of the absorbing gas.

## 3. LY$\alpha$ ABSORPTION LINES

The density of the neutral and ionized hydrogen, $n_{HI}$ and $n_{HII}$, can be calculated by balancing collisional ionization and photoionization with recombination:

$$\frac{n_{HII}}{n_{HI}} = 6.1 \times 10^3 \left(\frac{4}{1+\alpha}\right) \left(\frac{T}{10^4 \mathrm{K}}\right)^{3/4} \frac{J_{21}}{n_{-3}} + 1.9 \times 10^4 \left(\frac{T}{10^4 \mathrm{K}}\right)^{5/4} \exp\left(-\frac{158000}{T}\right) \quad (1)$$

where we have taken the various rates from Seaton (1960) and Osterbrock (1989), and $n_{-3} = n_{HII}/10^{-3}$ cm$^{-3}$. The background UV flux is assumed to be $J_\nu = J_0(\nu_1/\nu)^\alpha$, and $J_0 = 10^{-21}J_{21}$ erg/s/cm$^2$/Hz/sr is the flux at Lyman limit $\nu_1$; we take $\alpha = 1$ below. The recombination time at $r_c$ is about $5 \times 10^8$ yr, so the cooling calculations may be valid at redshift $z \lesssim 5 - 6$ for the clouds. In figure 1, we show $n_{HI}$ for $J_{21} = 1$ and $J_{21} = 0.01$. The peak of $n_{HI}$ at $r \simeq 4$ kpc for $J_{21} = 1$ is where the photoionization becomes dominant, and the increase of $n_{HI}$ at large radius $r$ is due to the enhanced recombination as $T$ drops.

The optical depth at frequency $\nu$ by neutral hydrogen seen at impact parameter $\rho$ is

$$\tau_\nu = \frac{\sigma_i}{\sqrt{\pi}} \int_{-\infty}^{\infty} n_{HI}(r) \exp\left[-\frac{(\nu - \nu')^2}{\Delta^2}\right] dy, \quad (2)$$

where $\sigma_i$ is the Ly$\alpha$ absorption cross-section, $y = (r^2 - \rho^2)^{1/2}$ is the line element along the line of sight (LOS), $\nu' = \nu_0(1 - yv/rc)/(1+z)$, $\nu_0$ is the rest-frame Ly$\alpha$ frequency, $\Delta = \nu'b/c$, and $b = (2kT/m_p)^{1/2}$. In figure 2, we show the Ly$\alpha$ line profiles for $\rho = 2, 3$, and 4 kpc for $J_{21} = 1$ (corresponding roughly to that inferred at $z \simeq 2 - 3$; e.g., Bechtold 1994). The contribution to the absorption in these cases comes mostly from the neutral hydrogen within $r_c$. The double-horn profiles are primarily caused by the peak in $n_{HI}$ at $r \simeq 4$ kpc. As $\rho$ exceeds $r_c$, the absorption becomes dominated by the cooled clouds.

If the LOS intercepts no clouds, the resulting line profile is a broad trough with a $b$-value comparable to the flow velocity ($\sim 170$ km/s), shown in figure 2 for $J_{21} = 1$ at $\rho = 20$ kpc. If the LOS intercepts a single cloud that dominates the absorption, one sees a narrow line with a width characteristic of the internal temperature $2 \times 10^4$ K. We show such an example in figure 2, assuming that a cloud with 4 times the ambient density and a radius of 1 kpc is intercepted at $y = 0$. The cloud size adopted is consistent with that for stable clouds based on the considerations of various cloud instabilities in gaseous halos (Mo 1995). In practice, clouds at different $y$ should also cause absorption, resulting in multiple lines with different strengths. However, the lines are likely to be dominated by clouds at $y = 0$ (moving perpendicular to the LOS) because the gas density is the highest here. This implies that the velocity difference in two LOS (e.g., toward a pair of QSOs) through the same galaxy should be small, consistent with recent observations (Dinshaw et al. 1994; (Bechtold et al. 1994; Dinshaw et al. 1994).

The cloud covering factor can be constrained from mass conservation: $f \lesssim (3/4\pi)(n/n_c)(l/r_c)$, where $n$ is the average flow density, $n_c$ and $r_c$ are the cloud density and radius, and $l$ is the path length of the LOS through the outflow. For the above cloud $f \lesssim 1$ if $l \sim \rho \sim 20$ kpc. The upper limit to the total mass in the clouds is just the total mass in the outflow, $M_{\text{total}}$. For figure 1, we have $M_{\text{total}} \simeq 5 \times 10^7$ and $2 \times 10^8$ M$_\odot$ within 20 and 100 kpc respectively, assuming that the outflow has not been interrupted; if the outflow has lasted only for a fraction of the cloud travel time $M_{\text{total}}$ is maller. The number of the absorption lines per unit neutral hydrogen column density can be estimated from $f(N_{HI}) \propto \rho/(dN_{HI}/d\rho)$, shown in figure 3. It is cut off at $N_{HI} \simeq 10^{14.3}$ cm$^{-2}$ since this is the maximum neutral column in this particular example. The calculated $f(N_{HI})$ is consistent with the observed distribution (e.g., Murdoch et al. 1986; Lu et al. 1991; Rauch et al. 1992).

## 4. EVOLUTION OF DWARFS

Assuming that each dwarf galaxy with outflow produces one strong (large equivalent width) Ly$\alpha$ absorption line, the number of such lines per unit $z$ by dwarfs of comoving density $n_0$ is

$$\frac{dN_c}{dz} = \frac{c\sigma n_0}{H_0} \frac{(1+z)}{(1+2qz)^{1/2}}, \qquad (3)$$

where $q$ is the deceleration parameter, and we have assumed that outflows last long enough to be seen in the absorption lines. The effective cross-section of the outflow is $\sigma = f\pi R^2$, where $f$ is the cloud covering factor and $R$ is the radius of the outflow regions. For figure 1, we have $R = 20$ kpc for $N_{HI} \gtrsim 10^{13}$ cm$^{-2}$. At large $r$ where $T$ is low, $n_{HI} \propto n_{HII}^2/J_{21} \propto r^{-4}/J_{21}$. Consequently, $\sigma \propto J_{21}^{-2/3}$ for a given $N_{HI}$. Thus the absorption lines shown in figure 2 at $\rho = 20$ kpc for $J_{21} = 1$ is almost identical to those at $\rho = 95$ kpc for $J_{21} = 0.01$ (confirmed by the detailed calculations).

Various estimates give $J_{21} \sim 0.01$ at $z \simeq 0$ (e.g., Kulkarni and Fall 1993; Dove & Shull 1994). Thus if we take $f = 1$ and $R = 100$ kpc, and compare with the observed $dN_c/dz \simeq 20$ at $z \sim 0$ (Bahcall et al. 1993), we derive $n_0 \simeq 0.2$ Mpc$^{-3}$. The number density of the local bright galaxies is about 0.02 Mpc$^{-3}$ (e.g., Loveday et al. 1992), implying a factor of $\sim 10$ more dwarf galaxies. (If $\lambda$ is larger than the assumed, $R$ is larger so $n_0$ required would be smaller). In the local group there are at least 27 dwarfs associated with the Galaxy and M31 within about 1 Mpc, and at least additional 16 within about a few Mpc (van den Bergh 1994). Clearly there are enough dwarfs locally to cause significant absorption lines. The fact that most dwarf galaxies in the local group are gas-poor (van den Bergh 1992) is consistent with that they have experienced significant gas losses. We note however that some dwarfs with extended HI disks also indicate that not all the dwarfs have lost their gas completely.

At large $z$, we assume $J_{21} \propto (1+z)^j$, so $dN_c/dz \propto n_0(1+z)^{\gamma_n}$ where $\gamma_n = -2j/3 + 1$ ($-2j/3 + 1/2$) for $q = 0$ (0.5). If we take $j \sim 3$ at $z \lesssim 2$ and $j \sim 0$ at $z \gtrsim 2$ as estimated from the QSO luminosity function (e.g., Madau 1992), from the observed $dN_c/dz \propto (1+z)^{0.8}$ at $z \lesssim 2$ (e.g., Morris et al. 1991; Bahcall et al 1993) and $dN_c/dz \propto (1+z)^{2.5}$ at $z \gtrsim 2$ (e.g., Murdoch et al. 1986; Lu et al. 1991; Rauch et al. 1992), we infer $n_0 \propto (1+z)^{1.8-2.3}$ at $z \lesssim 2$ and $(1+z)^{1.5-2.0}$ at $z \gtrsim 2$ for $q = 0 - 0.5$. Thus one expects a strong evolution for the number of the absorbing dwarfs if they are responsible for most of the Ly$\alpha$ absorptions. For example, $n_0$ is increased by about a factor of 2 from $z = 0$ to 0.4. Interestingly, deep extragalactic surveys show that the number of galaxies at apparent magnitude $b_J \sim 24$ in the blue band is larger than that expected from the local luminosity function by a similar factor (e.g., Maddox et al. 1990), while spectroscopic observations of these faint galaxies show that they have a median $z$ of 0.4, and most of them are sub-$L_*$ galaxies experiencing active star formation (e.g., Colless et al. 1993).

5. DISCUSSION

The implied rapid increase for the number of dwarfs toward large $z$ is therefore consistent with their experiencing higher rates of star formation in the past. This would naturally account for both the faint blue galaxies in the deep surveys and the Ly$\alpha$ forests. The onset of star formation inevitably leads to SN explosions and the production of a hot gas, which outflows from galaxies of small masses. Once the gas has become exhausted, star formation subsides, so we expect that the number of star-forming dwarfs is a decreasing function of time. The above then suggests that the absorption lines can provide a valuable tool for studies of redshift distribution of very faint galaxies. For example, our estimate in §4 gives that $n_0$ at $z \simeq 2$ is an order of magnitude larger than at $z \simeq 0$. In comparison, deep galaxy surveys show that the number of the faint galaxies at $b_J \sim 27$ is also an order of magnitude larger than expected. We thus suggest that most of galaxies at $b_J \sim 27$ may be at $z \sim 2$.

The energy injection rate into the outflow is $\dot{E} = 2c_i^2 \lambda = 5 \times 10^{39}$ erg/s, where $c_i$ is the sound speed at $r_i$ (Wang 1995) and the last equality follows from figure 1. In comparison, the energy rate from SN for a star formation rate $s_1 M_\odot$/yr with a Miller-Scale IMF is $2 \times 10^{41} s_1$ erg/s if each SN outputs $10^{51}$ erg (Wang & Silk 1993). Thus $\dot{E}$ above is equivalent to $s_1 = 0.03$. In practice, the efficiency of the SN heating is less than 100% so the implied $s_1$ should be higher, although the efficiency is uncertain, depending on the conditions of the interstellar gas (e.g., Larson 1974). The small star formation rate required in the dwarfs may bring difficulties to optical searches for them (e.g., Morris et al. 1993; Stocke et al. 1995). (Recall for the SMC $M_{B_T} = -16$ with $s_1 = 0.05$; Kennicutt et al. 1995). Moreover, for the reasons given above the star formation may have lasted only for a finite time (or sporadic), so dimming of the brightness by a few magnitudes from the peak is expected (e.g., Leitherer & Heckman 1995). Because the cooled clouds are unbound to galaxies, they can be at extremely large distances away from the dwarfs, so surveys encompassing galaxies at large distances are necessary. In the outflow models, the strength of the lines should decrease with the distance.

Surveys of the local gas-rich dwarf galaxies show that they are clustered around bright galaxies, although the cross-correlation is smaller than the auto-correlation among bright galaxies themselves (e.g., Alimi 1994). This is consistent with the studies of the cross-correlation between the Ly$\alpha$ absorbers and the bright galaxies (Morris et al. 1993; Lanzetta et al. 1995). Thus our model attributes the Ly$\alpha$ absorption clouds identified with bright galaxies (such as those by Lanzetta et al.) to dwarfs that are spatially correlated with (but unbound to) the bright galaxies. At high $z$, Ly$\alpha$ absorption lines show no large-scale clustering but on smaller scales (200-400 km/s) the clustering is suggested (e.g., Crotts 1989; Barcons & Webb 1991; Press & Rybicki 1993). In the outflow models, clustering on scales of a few 100 km/s is indeed expected, but the detailed (area-weighted) correlation depends on the number of clouds formed and the distribution of different dwarfs, hence requiring a careful modeling. In summary, in our "unified" picture of quasar absorption line systems, while the dwarfs that are bound to normal galaxies cause the high column (i.e., the heavy element and the Lyman limit) absorption lines (Wang 1993), the dwarfs that are unbound to but spatially correlated with the normal galaxies produce the low column Ly$\alpha$ clouds.

Fransson and Epstein (1982) first suggested that *galactic winds* may be responsible for the Ly$\alpha$ forests. However, they followed the motions of the gas with an approximation, which in fact can be shown to be not an exact solution for the outflows. More importantly, they did not include radiative cooling, which is crucial in accounting for the observed Ly$\alpha$ line profiles. Without radiative cooling, the absorption lines would be much broader than the observed ones (median $b \sim 35$ km/s; e.g., Rauch et al. 1992). Indeed, the only narrow line comparable to the observed Ly$\alpha$ from their wind model is the one resulting from the LOS passing through near the galactic center; all the other LOS result in broad lines with $b$-values characteristic of the initial sound speed ($\sim 100$ km/s) (see their figure 3). As a consequence, the cross-section for the narrow lines would be very small, so winds without radiative cooling cannot contribute significantly to the Ly$\alpha$ forests. Rees (1986), and more recently Miralda-Escude & Rees (1993) and Cen et al. (1994), considered the Ly$\alpha$ clouds as a result of structure formation on small mass scales ($v_{cir} \lesssim 30$ km/s) where gas cannot collapse to form stars in the context of the CDM model. We wish to point out here however that structures on larger scales (where stars form) can naturally account for various characteristics of the Ly$\alpha$ forests and in particular the heavy elements recently detected in them (Tytler 1995).

The author benefited from discussions with Annette Ferguson, Harry Ferguson, Gerhardt Meurer, Hou Jun Mo, Dave Bowen, and Ray Weymann. Support from Alan C. Davis Fellowship at the Johns Hopkins University and STScI is gratefully acknowledged.

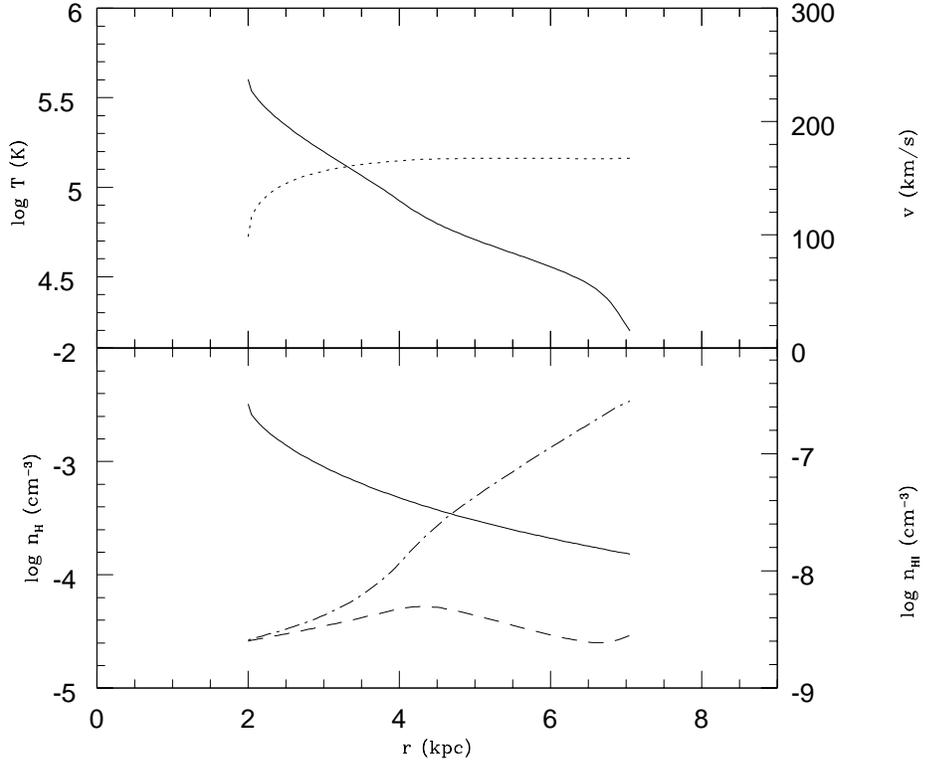

Fig. 1

**Fig. 1.** Outflow variables as a function of the radius. Top panel: the gas temperature (solid curve; left ordinate) and the flow velocity (dotted; right ordinate). Bottom panel: the total hydrogen density (solid, left ordinate), and the neutral hydrogen density (right ordinate) for the UV ionizing flux $J_{21} = 1$ (dashed) and $J_{21} = 0.01$ (dash-dotted).

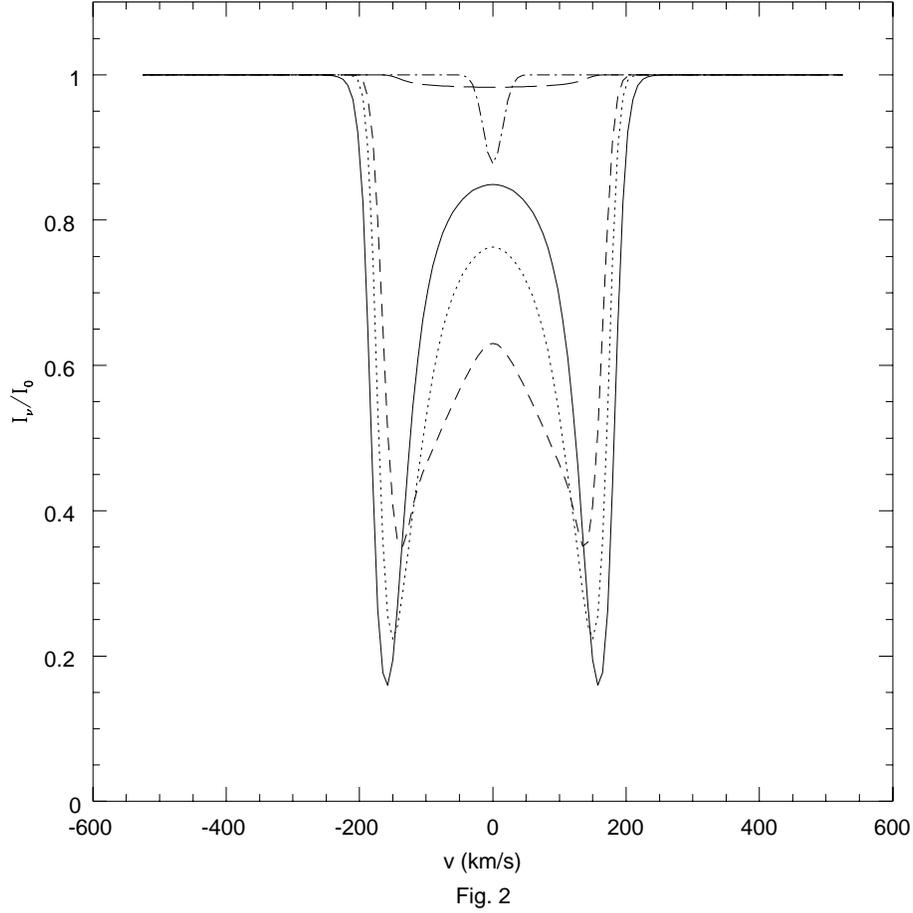

**Fig. 2.** The Ly$\alpha$ absorption line profiles, $I_\nu/I_0 = e^{-\tau_\nu}$, as a function of the velocity shift $v \equiv c\delta\nu$ for $J_{21} = 1$. The impact parameter to the line of sight is $\rho = 2$ (solid), 3 (dotted), 4 (dashed) kpc, respectively. For $\rho = 20$ kpc, we have assumed that the line of sight intercepts no clouds but only diffuse gas (long-dashed) or intercepts a single cloud with a radius of 1 kpc and an overdensity of 4 (dash-dotted).

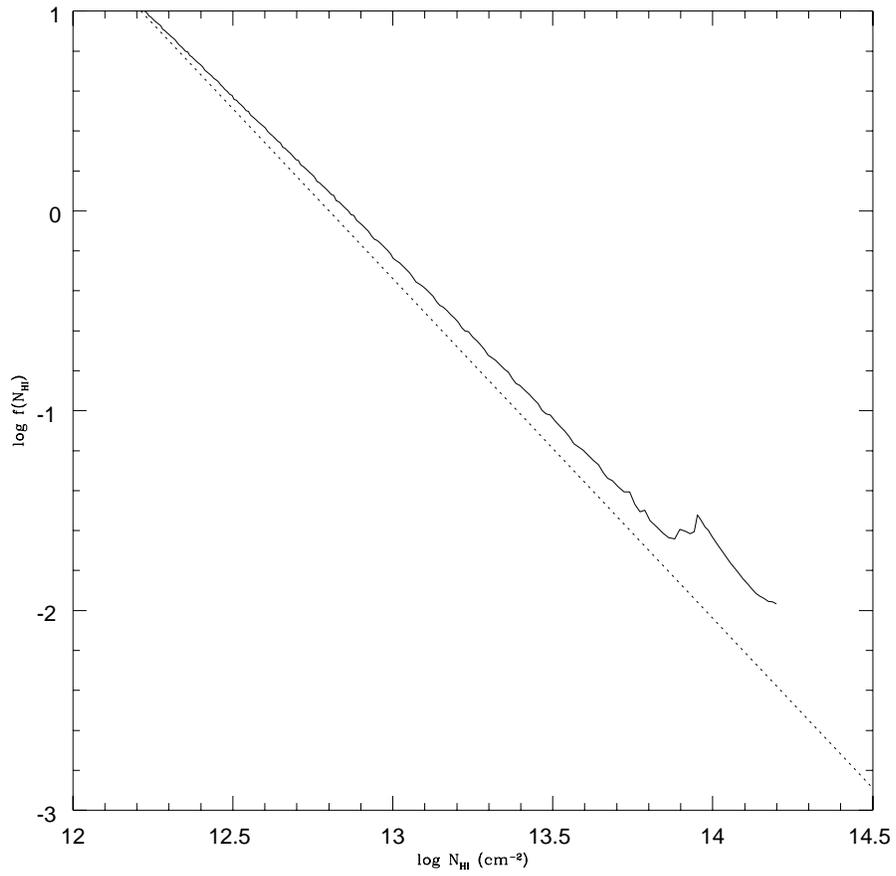

Fig. 3

**Fig. 3.** The distribution of the neutral hydrogen column density (solid) (arbitrarily normalized). The dotted line (not a fitting) is a power law $N_{HI}^{-1.7}$.